# Plausibility of Quantum Coherent States in Biological Systems

[1,2,3]V. Salari, [4,5]J. Tuszynski, [6,7]M. Rahnama, [8]G. Bernroider

[1]Institut de Mineralogie et de Physique des Milieux Condenses, Universite Pierre et Marie Curie-Paris 6, CNRS UMR7590, France
[2]Boite courrier 115, 4 place Jussieu, 75252 Paris cedex 05, France
[3] BPC Signal, 15 rue Vauquelin, 75005 Paris, France
[4]Department of Experimental Oncology, Cross Cancer Institute, 11560 University Avenue Edmonton, AB T6G 1Z2, Canada
[5]Department of Physics, University of Alberta, Edmonton, AB Canada
[6]S. B. University of Kerman, Kerman, Iran
[7]Afzal Research Institute, Kerman, Iran
[8]Department of Organismic Biology, University of Salzburg, Hellbrunnerstrasse 34, Salzburg, Austria

**Abstract:**

In this paper we briefly discuss the necessity of using quantum mechanics as a fundamental theory applicable to some key functional aspects of biological systems. This is especially relevant to three important parts of a neuron in the human brain, namely the cell membrane, microtubules (MT) and ion channels. We argue that the recently published papers criticizing the use of quantum theory in these systems are not convincing.

### 1) Quantum theory and biological systems

Biological systems operate within the framework of irreversible thermodynamics and nonlinear kinetic theory of open systems, both of which are based on the principles of non-equilibrium statistical mechanics. The search for physically-based fundamental models in biology that can provide a conceptual bridge between the chemical organization of living organisms and the phenomenal states of life and experience has generated a vigorous and so far inconclusive debate [1,2]. Recently published experimental evidence has provided support for the hypothesis that biological systems use some type of quantum coherence in their functions. The nearly 100% efficient excitation energy transfer in photosynthesis is an excellent example [3]. Living systems are composed of molecules and atoms, and the most advanced physical theory describing interactions between atoms and molecules is quantum mechanics. For example, making and breaking of chemical bonds, absorbance of frequency specific radiation (e.g. in photosynthesis and vision), conversion of chemical energy into mechanical motion (e.g. ATP cleavage) and single electron transfer through biological polymers (e.g. in DNA or proteins) are all quantum effects. Regarding the efficient functioning of biological systems, the relevant question to ask is how can a biological system with billions of semi-autonomous components function effectively and coherently? Why providing a complete explanation remains a major challenge, quantum coherence is a plausible mechanism responsible for the efficiency and co-ordination exhibited by biological systems [4].

The hypothesis invoking long-range coherence in biological systems was proposed by H. Fröhlich's [5-7] and followed by detailed investigations by Tuszynski et al. [8-22], Pokorny [23-25], Mesquita et al. [26-28] and others for over three decades. The possible role played by coherent states manifested outside low temperature physics has attracted considerable interest in both the physics and biology communities. Despite the potential power of quantum mechanics to explain coherent phenomena, there are serious challenges involved in applying it in the context of a living system. For instance, in order to have a very high degree of coherence between bio-molecules, Bose-Einstein condensation may be a viable mechanism, but we note that the ambient temperature in a living system is likely to be too high for this phenomenon to occur. Also, the sizes of bio-molecules are very large by physical standards in order to be regarded as typical quantum systems. Moreover, because of the noisy environment, according to decoherence theory, quantum states of these mesoscopic bio-molecules would collapse very rapidly. However, remarkably there is no obvious limitation placed on the Schrodinger equation for its use only in atomic-scale systems. It is a universal equation and it can even be used for the entire universe (as is the case with quantum gravity applications). The boundary between quantum theory and classical physics is still largely unknown. Quantum theory obviously applies on length scales smaller than atomic radii but beyond that it is not entirely clear where it should be superseded by Newtonian mechanics. Superconductors, lasers, superfluids, semiconductors etc., are examples of macroscopic-scale physical systems that behave quantum mechanically, so it is also possible that biological systems operate based on quantum principles at least in some of their functions. Here we argue that recent criticisms of the use of quantum mechanics in biology are not very convincing since they ignore the already existing evidence for quantum effects in biological systems. In this paper we investigate three particular systems of special importance: the cell membrane, microtubules (MTs), and ion channels which are some of the most important parts of a neuron in the human brain. We argue that these subsystems are the best candidates for possible sites of quantum effects.

2) **A brief overview of the criticism of coherence and decoherence**

**2-1) Membrane**

The original Fröhlich model was general and did not limit the mechanism of biological coherence to any particular cellular structure. In his model, when the energy supply exceeds a critical level, the dipolar ensemble of biologically relevant molecules populates a steady state of non-linear vibrations characterized by a high degree of structural and functional order [29]. This (electrically polarized) ordered state expresses itself in terms of long-range phase correlations, which are physically similar to such phenomena as superconductivity and superfluidity, where the behaviour of particles is collective and inseparable. The existence of very strong static electric fields across the cell membrane led Fröhlich to consider cellular membranes to be the source of the postulated coherent vibrations [29].

Froehlich considered a model for biological systems consisting of three parts:

1) A system of oscillators
2) A heat bath
3) An external source which pumps energy into the system incoherently.

Froehlich [5] derived a formula for the net rate of energy loss $L_i^1$ of the mode with frequency $\omega_i$ (containing $n_i$ quanta) which can be written as

$$L_i^1 = \phi(T)(n_i e^{\hbar\omega_i/kT} - 1 - n_i) \qquad (1)$$

where $\hbar$ is Planck's constant ($\frac{h}{2\pi}$), $k$ is Boltzmann's constant, $T$ is the temperature of the system and the heat bath, and $\phi(T)$ is a function of $T$. Recently, Reimers et al. [2] have shown that a very fragile Froehlich coherent state may occur at sufficiently high temperatures and concluded that there is no possibility for the existence of Froehlich coherent states in biological systems. However, it should be noted that in equation (1) they replaced the function $\phi(T)$ with a constant $\varphi$ at 0 Kelvin. This replacement causes equation (1) to be plagued by serious problems in the limit

$$T \to 0, \quad \frac{\hbar\omega_i}{kT} \to \infty, \quad e^{\frac{\hbar\omega_i}{kT}} \to \infty \quad \Rightarrow \quad L_i^1 \to \infty \qquad (2)$$

It means that the net rate of energy loss tends to infinity near the absolute zero, so it makes this assumption and its consequences unphysical. In the second order considered by these authors, in the case of a three body system, the same problem emerges, too. Also they have demonstrated several diagrams in terms of effective temperature which was defined by themselves as $T_{eff} = \frac{T_S}{T}$, where $T_S$ is the temperature of system and $T$ is the temperature of the thermal bath. Their effective temperature relation is

$$\frac{T_S}{T} \approx 1 + \frac{s}{\phi} e^{-\frac{\hbar\omega_0}{2kT}} \qquad (3)$$

where $s$, $\phi$ and $\omega_0$ are constants. Here $T_{eff}$ is not a well-defined parameter because on both sides of the relation in (3) we have temperature of the heat bath $T$. Therefore, in the limit, both sides of the relation cannot be reconciled with each other:

$$T \to 0, \quad -\frac{\hbar\omega_0}{2kT} \to -\infty, \quad e^{-\frac{\hbar\omega_0}{2kT}} \to 0 \quad thus \quad T_{eff} \to \infty \quad while \quad (1 + \frac{s}{\varphi} e^{-\frac{\hbar\omega_0}{2kT}}) \to 1 \qquad (4)$$

$$T \to \infty, \quad -\frac{\hbar\omega_0}{2kT} \to 0, \quad e^{-\frac{\hbar\omega_0}{2kT}} \to 1 \quad thus \quad T_{eff} \to 0 \quad while \quad (1 + \frac{s}{\varphi} e^{-\frac{\hbar\omega_0}{2kT}}) \to 1 + \frac{s}{\varphi} \qquad (5)$$

They have used the effective temperature parameter for the Wu-Austin Hamiltonian [30-32] and considered it in the high temperature limit based on the references numbered [49, 50] and

[51] in their paper while there are actually only 49 references cited in this paper. Their diagrams are mostly based on the effective temperature parameter and hence are, in our opinion, not acceptable due to the internally contradictory argument used in their derivations.

### 2-2) Microtubules

The criticism raised by Reimers et al. [2, 37] is mainly directed against the so-called Orch OR model which was proposed by Penrose and Hameroff to introduce a physical basis for consciousness. In some formulations of the OrchOR model, a manifestation of quantum coherence involved the involvement of Froehlich coherent states in MTs [33-35]. MTs are highly ordered in the neurons of the brain and can indeed be regarded as good candidates for supporting Froehlich coherent states. In this context, the conclusions of section 2 apply to MTs as well. Therefore, we believe that it is still hypothetically possible to generate Froehlich coherent states in MTs. However, another issue that arises when considering quantum states for MTs is the rapid *decoherence* problem. The question is "how is it possible for MTs to be in a coherent state while the environment surrounding them is relatively hot, wet and noisy?" According to the decoherence theory, while macroscopic objects obey quantum mechanics, their interactions with the environment cause decoherence, which destroys quantum effects [36]. For macroscopic particles there are two main 'natural' ways of experiencing decoherence, first is due to collisions with other particles and the second is the thermal emission of radiation due to the internal heat of an object [38, 41]. In the latter case, the decoherence time of the system is given by

$$T_{dec} = \frac{1}{10^{36} a^6 T^9 (\Delta x)^2} \qquad (6)$$

where $a$ is the size of the object (diameter or length), $T$ is the absolute temperature and $\Delta x$ is the superposition distance of the object [38]. At room temperature $T=300K$ and the size $a=8nm$ for the tubulin dimer (i.e. the elementary constituent of an MT) and considering $\Delta x = 8nm$ for its superposition distance, the decoherence time would be on the order of $T_{dec} \sim 10^6 s$. This indicates that thermal photons cannot cause decoherence in this case. Hence we concentrate on the first case which is decoherence due to the scattering by environmental particles. Tegmark [39] has calculated decoherence times for MTs based on the scattering between MTs and environmental particles. Based on the collisions of ions with MTs, he has obtained the decoherence times on the order of:

$$\tau = \frac{D^2 \sqrt{mkT}}{Ngq^2} \approx 10^{-13} s \qquad (7)$$

where $D$ is the tubulin diameter, $m$ is the mass of the ion, $k$ is Boltzmann's constant, $T$ is temperature, $N$ is the number of elementary charges in the microtubule interacting system, $g = \frac{1}{4\pi\varepsilon_0}$ is the Coulomb constant and $q$ is the charge of an electron. Hagan et al. [40] have shown that Tegmark used wrong assumptions for his investigation of MTs. Another main objection to the estimate in equation (7) is that Tegmark's formula yields decoherence times that increase with temperature contrary to a well-established physical experience and the observed behavior of quantum coherent states. In view of these (and other) problems in Tegmark's estimates, Hagan et al. [40] assert that the values of quantities in Tegmark's relation are not correct and thus the decoherence time should be approximately $10^{10}$ times larger leading to a ms range of values for typical decoherence times. According to Hagan et al., MTs in neurons could avoid decoherence via several mechanisms over sufficiently long times for quantum processing to occur there.

**2-3)    Ion Channel and Selectivity Filter (SF)**

Ion channels are proteins in the membranes of excitable cells that cooperate for the onset and propagation of electrical signals across membranes by providing a highly selective conduction of charges bound to ions through a channel like structure. The SF is a part of the protein forming a narrow tunnel inside the ion channel which is responsible for the selection process and fast conduction of ions across the membrane. The determination of atomic resolution structure of the ion channel and selectivity filter by Mac Kinnon led to the award of the Nobel prize for chemistry in 2003 [42]. The SF is a very narrow region of the protein. In the potassium selective KcsA bacterial channel, that frequently serves as a model structure, this regions extends to about 1.2 nm length and 0.3 nm in diameter. At the atomic scale the filter region exposes negative charges owing to the lone electron pairs from oxygen ions bound to 20 carbonyl groups arranged into 5 rings of the lining 'P-loop' peptide. Alltogether this structure provides a highly ordered atomic coordination pattern among oxygens and ions. If a positive charged ion, such as sodium or potassium enters the SF, the ion can be transiently trapped by Coulombic interactions with the negative charges provided by the oxygen ions. The physical action orders associated with the transient trapping of ions from the macroscopic scale down to the quantum scale have been analysed previously [43]. It has been concluded [43-46] that quantum theory is needed to explain the states of ions in the selectivity filter and hence the function of an ion channel. In contrast, Tegmark has calculated decoherence times for the superposition of ions crossing the entire membrane based on a simple ion-pore diffusion model [39]. The calculated decoherence times for crossing ions in this work are derived from the scattering with environmental particles based on the Coulomb interaction between ions and particles. He has assumed that ions are in a superposition state of inside and outside of the cell and are separated by a distance of 10 nm. In the view of atomic scaled resolution maps and recent molecular dynamics studies of the filter region in this protein, this type of interaction is oversimplified and the pore-diffusionscattering model is not applicable to describe ion protein interactions. Further, Tegmark introduced a function for the decoherence rate [47] which is composed of two parts: one for short wavelengths and the other for long wavelengths. Every scattering calculation based on the Coulomb interaction and Tegmark's decoherence rate function leads to decoherence times being directly proportional to temperature according to relations such as $\tau_{dec} \propto \sqrt{T}$, $\tau_{dec} \propto \sqrt{T^2}$, $\tau_{dec} \propto \sqrt{T^3}$, etc. Therefore, it can be expected that subsequent calculations based on these criteria are flawed in the high-temperature limit, i.e. as temperature approaches infinity, decoherence time increases too, and if temperature approaches absolute zero decoherence time approaches zero, a very unphysical situation. In contrast to Tegmarks conclusions, we suggest that his formulation of the problem cannot apply to the ion-protein coordinated states and his calculations do not address the temperature dependence of decoherence times correctly. Instead it is obvious that this problem remains unresolved and there is no simple and general relation between decoherence time and temperature in ion channels.

Based on a number of physical examples, we may expect biological coherence to occur under special circumstances at physiological temperatures. For example, it is well-known that lasers can maintain their coherence at high temperatures due to external pumping. Moreover, quantum spin transfer between quantum dots connected by benzene rings (the same structures that is found in aromatic hydrophobic amino acids) is more efficient at warm temperature than at absolute zero [48]. Further, a simple calculation of the translocation time of ions through an ion channel if based on a particle point of view, is quite different from what is obtained from experimental data based on X-ray crystallography and MD simulations [49-54]. The difference in the estimates of these translocation times can involve a factor up to $10^2$ to $10^5$ which indicates that the previous calculations using the particle point of view (i.e. classical physics) should be corrected when applied to the SF of channel proteins. Ions coordinated by the filter carbonyls should be considered as wave-packets rather than particles and their translocation as well as filter states must involve some interferences that account for the observed discrepancies between classical estimations and real observations. Recently, some of the expected consequences behind quantum interferences in the filter region of ion channels have been introduced into the classical equations of motion that lead to Hodgkin-Huxley type membrane potentials. As a result, these 'semi-classical' equations of ion motion are shown to carry the signatures of quantum-interferences [45] and can predict signal onset characteristics that are in accord with real recordings from central brain neurons [55].

### 3) Conclusion:

We have argued above that the objections raised against the feasibility and role of quantum effects in biological systems are not tenable. It is still an open question as to how a macroscopic object is classical while its constituent atoms and molecules are quantum mechanical in nature. As atomic-scale quantum systems compose into large molecules they become classical objects. In biology however, the challenging question is whether and how the initial quantum properties extend into the 'functional domain' of the emerging classical systems. In the case of energy transfer in photosynthesis, magnetic compass sensing with receptor proteins, microtubules and ion conduction in channel proteins, transient quantum coherences may well play a decisive role to explain the observed functional states of classical molecular systems. However, it seems that decoherence theory has not solved this problem, and hence we pose an important question: "What is the meaning of classicality when a large or complex system (as a quantum system) collapses to become a classical entity while the components (atoms or molecules) are still quantum mechanical?". At least in biological systems, we can expect that the emergence of classicality will involve some quantum signatures that cannot be ignored in functional explanations. Answers along this way will most probably play an increasing role for the understanding of organizational complexity and functions in living systems.

**Acknowledgments:**

Research support from NSERC (Canada) awarded to J.A.T. is gratefully acknowledged. Vahid Salari acknowledges BPC Signal for their support. We thank Roman Fuchs from the Neurosignaling Unit, Univ. Salzburg, Austria for kindly assisting in preparing the manuscript.